\documentstyle[epsf]{l-aa}
\newcommand{\ms}{M$_{\odot}$}
\newcommand{\teff}{$T_{\rm eff}$}
\newcommand{\lgg}{log\,$g$}
\newcommand{\kms}{km\,s$^{-1}$}
\newcommand{\vrad}{$v_{\rm rad}$}
\newcommand{\ea}{et~al.~}
\newcommand{\w}{$\varpi$}
\newcommand{\mua}{$\mu_{\alpha}{\rm cos}\delta$}

\newcommand{\dmuac}{$\delta(\mu_{\alpha}{\rm cos}\delta)$}
\newcommand{\dmub}{$\delta(\mu_{\alpha}{\rm .c})$}
\newcommand{\mud}{$\mu_{\delta}$}

\newcommand{\dmud}{$\delta(\mu_{\delta})$}

\begin{document}
  
\title{Hot subdwarf stars: 
  galactic orbits and distribution perpendicular to the plane}

\author{K.S. de Boer\inst{1}
\and Y. Aguilar S\'anchez\inst{1}
\and M. Altmann\inst{1}
\and M. Geffert\inst{1}
\and M. Odenkirchen\inst{1,2}
\and J.H.K. Schmidt\inst{1}
\and J. Colin\inst{2}
}

\institute{Sternwarte, Univ. Bonn, Auf dem H\"ugel 71, D-53121 Bonn, Germany
\and
Observatoire de Bordeaux, CNRS/INSU, F-33270 Floirac, France
}

\date{Received: 24 April 1997 / Accepted 24.6.97}

\thesaurus{05.01.1, 08.08.2, 08.11.1, 08.16.3, 10.08.1, 10.19.3}

\offprints{deboer@astro.uni-bonn.de}

\maketitle

\markboth{K.S. de Boer et al., Hot subdwarfs and galactic orbits}
	{K.S. de Boer et al., Hot subdwarfs and galactic orbits}

\begin{abstract}

The spatial distribution and the population nature 
of subdwarf B type stars in the galaxy is 
investigated based on the kinematics of these stars. 
With new and available proper motions, radial velocities, and distances, 
the orbits of 41 stars have been calculated using a galactic mass model. 
The orbits are well behaved and 10 stars reach to $|z| \geq 2$ kpc.
Many orbits are very eccentric, reaching in to just 2 kpc from the 
galactic centre, or veering out to beyond 20 kpc. 
None of the stars can be identified uniquely 
as classical Population II objects. 

The average eccentricity $ecc$ of the orbits of our sample is 0.24, 
the average normalised $z$-extent $nze$  of the orbits is 0.16, and  
the asymmetric drift of our sample is $-36$ \kms. 
This suggests that our sample of sdB stars is part of a population of 
thick disk stars. 

A statistical analysis of the orbits shows that the sub\-dwarf stars 
have a spatial distribution in $z$ compatible with an exponential one 
with a scale height $h_z$ of about 1.0 kpc. 
However, since only few stars reach to large $z$ 
the spatial distribution is only well defined to $z \simeq 2$ kpc. 

The distribution in $z$ shows a relative minimum near $z=0$ pc
and has maxima near $z=300$ pc. 
This reflects the smaller probability to find the stars in the disk 
than away from the disk, as expected for any orbit reaching to larger $z$. 
Scale height studies based on limited samples of stars 
in specified directions 
can therefore easily be flawed when they do not reach to 
large enough distances to overcome this aspect of the $z$-distribution. 

\end{abstract}

\section{Introduction}

Hot subdwarfs are blue, horizontal-branch like stars, 
re\-pre\-sent\-ing the late stages of evolution of stars having 
started with less than about 2.5 \ms\ on the main sequence. 
The hotter subdwarf B (sdB) stars form a well defined group. 
Their luminosity is of the order of 10 L$_{\odot}$ and their 
surface temperatures are roughly between 2 10$^4$ and 3 10$^4$ K. 
The age of the sdB stars does not follow from the stellar properties. 
The abundance of metals in subdwarf star atmospheres 
is possibly affected by upward convection of processed material 
but more likely by gravitational downward diffusion of the heavy elements. 
Thus the normally low metal abundance seen in sdB stars is not 
an indicator for their age. 

Our studies of hot subdwarf stars have two main goals: 
one is to determine the parameters and the evolutionary state 
of the subdwarf stars in the framework of stellar evolution; 
the other is to investigate the structure of the Milky Way 
using these stars. 
A relation between these two goals can be found in the connection 
between the age of subdwarf stars and their spatial distribution. 
Old stars will more likely be distributed in a thicker disk, 
due to the heating up of their average kinematics 
through interactions with other stars. 
Young stars, on the other hand, are rather moving inside the thin 
gaseous galactic disk where they originated. 
Still, the present day location of a star is, 
in general, no indication for its age. 

In this paper we will investigate the kinematic behaviour 
exemplified by the orbit. 
The characteristics of the orbits may contain information about 
the formation epoch and place of the stars, 
information not accessible through the metal content.  
Our investigations will show whether the star is in its motion confined 
in the thin disk of the Milky Way, 
or that it may reach to large $z$-distances similar to objects 
of the halo population.

\begin{table*}
\caption[]{Observational data for the stars for which orbits are calculated}
\begin{tabular}{llllrrrrll}
\hline
Name$^a$ & \multicolumn{2}{l}{RA (Equinox 2000) DEC $^b$} & 
	spectral & $d$ & \vrad $^c$ &
	\mua & \mud & ref. star & ref. p.m.\\
	& hr, min, sec & \ \ \ \ $^\circ$ \ \ $\arcmin$ \ \ $\arcsec$ & 
	type & pc & \kms & mas/yr & mas/yr & \\
\hline 
PG 0004+133 & 00 07 33.770 & +13 35 57.65 & sdB & 1430 & $-$37 & 
	+3.0 & $-$25.0 & PII & T+97\\
PG 0039+049 & 00 42 06.110 & +05 09 23.37 & sdB & 1050 & +87 & 
	+7.5 & $-$12.0 & PII & T+97\\
CD $-$38 222 $^a$ & 00 42 58.263 & $-$38 07 36.97 & sdB & 325 & $-$1 & 
	+34  & $-$6\ \  & B+97, H+84 & PPM\\
PG 0101+039$^a$ & 01 04 21.670 & +04 13 37.26 & sdB & 450 & {\sc vs}+89 & 
	+12.2 & $-$40.0 & PII & T+97\\
PG 0133+114 & 01 36 26.259 & +11 39 30.95 & sdB & 770 & $-$77 & 
	+20.7 & $-$34.4 & PII & C+94\\
PHL 1079 & 01 38 26.93 & +03 39 38.0 & sdB & 810 & 0 & 
	11.1 & $-$17.9 & PII, PIV & Table\,2\\
PG 0142+148 & 01 45 39.57  &  +15 04 41.5 & sdB & 1170 & $-$131 & 
	$-$17.4 & $-$0.4 & PII & Table\,2\\
PG 0212+148 & 02 15 11.078  &  +15 00 04.55 & sdB & 1750 & +50 & 
	$-$3.8 & $-$9.2 & PII & Table\,2\\
PG 0212+143 & 02 15 41.602  & +14 29 17.97  & sdB & 1850 & +77 & 
	+11.2 & $-$1.4 & PII & Table\,2\\
PG 0242+132 & 02 45 38.855 & +13 26 02.41 & sdB & 1390 & +11 & 
	+17.2 & $-$9.7  & PII & Table\,2\\
PG 0856+121 & 08 59 02.723 & +11 56 24.73 & sdB & 990 & +85 & 
	$-$19.4 & $-$19.8 &  PII & C+94\\
PG 0907+123 & 09 10 07.6  & +12 08 26.1  & sdB & 1520 & {\sc vg}+85 & 
	+6.4 & $-$2.6 &  PII & C+94\\
PG 0918+029 & 09 21 28.230 & +02 46 02.25 & sdB & 1040 & {\sc vg} +3 & 
	$-$28.5 & $-$20.0 & PII & T+97\\
PG 0919+273$^a$ & 09 22 39.830 & +27 02 26.15 & sdB & 350 & $-$33 & 
	+22.9 & $-$19.8 & S+94 & K+87\\
PG 1101+249$^a$ & 11 04 31.731 & +24 39 44.75 & sdB & 390 & $-$48 &
	$-$30.3 & +16.0 &  S+94 & K+87\\
PG 1114+073$^a$ & 11 16 49.670 & +06 59 30.83 & sdB & 450 & +4 & 
	$-$12.3 & $-$14.4 & S+94 & K+87\\
PG 1232$-$136$^a$ & 12 35 18.915 & $-$13 55 09.31 & sdB & 570 & {\sc vg}+55 &
	$-$46.4 & $-$1.7 & S+94 & K+87\\
PG 1233+427$^a$ & 12 35 51.641 & +42 22 42.64 & sdB & 320 & +61 & 
	+3.6 & $-$18.1 & S+94 & K+87\\
PG 1256+278$^a$ & 12 59 21.266 & +27 34 05.22 & sdB & 780 & +64 & 
	$-$24.6 & +3.5 & S+94 &  K+87\\
PG 1343$-$101$^a$ & 13 46 08.069 & $-$10 26 48.27 & sdB & 720 & {\sc vg}+49 &
	$-$28.0 & $-$3.7 & S+94 & K+87\\
PG 1432+004 & 14 35 19.833 & +00 13 47.96 & sdB & 760 & +41 & 
	$-$9.4 & $-$25.8 & PII &  C+94\\
PG 1433+239$^a$ & 14 35 20.359 & +23 45 27.52 & sdB & 470 & $-$56 & 
	$-$3.5 & $-$18.5 & S+94 & K+87\\
PG 1452+198 & 14 54 39.810 & +19 37 00.88 & sdB & 810 & +51 & 
	$-$7.2 & $-$21.0 & PII & T+97\\
PG 1519+640 & 15 20 31.320 & +63 52 07.95 & sdB & 650 & {\sc vg}+37 & 
	+28.7 & +44.2 & PII &  T+97\\
PG 1619+522 & 16 20 38.740 & +52 06 08.78 & sdB & 770 & {\sc vg}$-$36 &
	$-$3.6 & +9.0 & PII & T+97\\
PG 1647+252$^a$ & 16 49 08.974 & +25 10 05.74 & sdB & 710 & +26 & 
	$-$3.8 & +12.3 & PIV & K+87\\
PG 1708+602 & 17 09 15.900 & +60 10 10.79 & sdB & 1790 &$-$8 & 
	$-$14.9 & +12.1 & PIV & T+97\\
PG 1710+490 & 17 12 18.740 & +48 58 35.88 & sdB & 720 & $-$44 & 
	+10.8 & $-$7.0 & PII, PIV & T+97\\ 
PG 1722+286 & 17 24 11.970 & +28 35 26.93 & sdB & 870 & $-$51 & 
	$-$4.0 & +10.0 & PIV & T+97\\
PG 1725+252 & 17 27 57.390 & +25 08 35.69 & sdB & 660 & {\sc vg}+22 &
	$-$17.7 & +9.0 & PIV & T+97\\
PG 1738+505 & 17 39 28.440 & +50 29 25.11 & sdB & 970 & +22 & 
	$-$7.6 & +9.0 & PIV & T+97\\
HD\ \ 205805  $^a$ & 21 39 10.699 & $-$46 05 51.35 & sdB & 265 & $-$57 &
	+79 & $-$16\ \  & B+97, S83 & PPM\\
PG 2204+035 & 22 07 16.490 & +03 42 19.82 & sdB & 1180 & +81 & 
	+7.5 & $-$6.0 & PII & T+97\\
PG 2218+020 & 22 21 24.83 & +02 16 18.6 & sdB & 1310 & 21 &
	+1.2 & $-$11.8 & PVII & Table\,2\\
PG 2226+094 & 22 28 58.41 & +09 37 21.8 & sdB & 1130 & 29 & 
	+14.3 &  +0.4 & PVII & Table\,2 \\
PG 2259+134 & 23 01 45.82 & +13 38 37.5 & sdB & 1550 & 16 & 
	+0.6 &  $-$9.6 & PIV & Table\,2 \\
Feige 108 $^a$ & 23 16 12.41 & $-$01 50 34.50 & sdB & 395 & 40 & 
	$-$7.2 & $-$18.4 & S+94 & Table\,2\\
Feige 109 $^a$ & 23 17 26.890 & +07 52 04.93 & sdB & 1130 & $-$19 & 
	+1.5 & +6.0 & PII & Table\,2\\
PG 2337+070 & 23 40 04.83 & +07 17 11.00 & sdB & 770 & $-$27 & 
	$-$19.0 & $-$38.8 & PVII & Table\,2\\
PG 2349+002 & 23 51 53.26 & +00 28 18.00 & sdB & 820 & $-$84 & 
	$-$14.6 & $-$19.3 & PVII & Table\,2\\
PG 2358+107 & 00 01 06.730 & +11 00 36.32 & sdB & 830 & $-$19 & 
	$-$3.0 & $-$14.0 & PII & T+97\\
\hline
\end{tabular}

\noindent
$^a$ other star names are:
CD\,$-$38\,222  = SB\,290;
PG\,0101+039   = FB\,13;
PG\,0919+273   = NPM\,+27\,0638;
PG\,1114+073   = NPM\,+07\,0956;
PG\,1232$-$136 = NPM\,$-$13\,1270;
PG\,1233+427   = NPM\,+42\,0772;
PG\,1256+278   = NPM\,+27\,1076; 
PG\,1343$-$101 = NPM\,$-$10\,1622;
PG\,1433+239   = NPM\,+23\,0716;
PG\,1647+252   = NPM\,+25\,0875;
HD\,205805     = FB\,178;
Feige\,108 = PG\,2313$-$021;
Feige\,109 = PG\,2314+076; star names NPM go back to K+87 (see below)\\
$^b$ positions are new (Table\,2) or from K+87, C+94, T+97, the PPM 
	(see last column); 
the Epochs of these data sets are: 
Table\,2, 1970; K+87, 1960; C+94, 1950; T+97, 1990.\\
$^c$ radial velocity was taken first from Papers II, IV, and VII, 
	or was determined by us from spectra in our data base; 
	then from further literature cited. 
	{\sc vg} or {\sc vs}: radial velocity variable according to 
	Green \ea (1997) or Saffer (1994)\\
B+97 = de Boer \ea (1997); 
C+94 = Colin \ea (1994); 
H+84 = Heber \ea (1984);
K+87 = Klemola \ea (1987); 
PII = Moehler \ea (1990); 
PIV = Theissen \ea (1993); 
PVII = Aguilar S\'anchez \ea (in prep.);
PPM = R\"oser \& Bastian (1991);
S83 = Stetson (1983);
S+94 = Saffer \ea (1994);
T+97 = Thejll \ea (1997)

\end{table*}

A first sample of data addressing the population nature of subdwarf stars 
has been presented by Colin \ea (1994). 
They showed that indeed most stars in their (small) sample had disk orbits 
but 1 star had an orbit with $z$-distance maxima ranging from 8 to 20 kpc.
Evidence for more stars with halo orbits was given by de Boer \ea (1995). 

We have investigated 41 stars for their kinematic behaviour. 
The choice of stars was solely determined by the availability 
of the parameters necessary to calculate orbits. 
The stars selected are listed in Table 1. 
The sample includes the stars already investigated by Colin \ea (1994). 

For 12 stars new absolute and accurate proper motions 
have been determined (see Sect.\,2.1) allowing to 
evaluate the difference between the various astrometric catalogues 
and improving on the data for some of the Colin \ea stars. 

After having presented the orbits we analyse their shapes 
and try to identify the oldest objects of the sample (Sect.\,3). 
The range of $z$-values the stars reach is then used to investigate 
the spatial distribution of sdB stars in the Milky Way (Sect.\,4). 

%
\section{The data}

For our study we base ourselves on new and published proper motions, 
radial velocities, and distances for subdwarf stars (Table 1). 
Most of the stars are part of a programme to investigate 
the nature and distance of horizontal-branch type stars 
(see Moehler \ea 1990, Theissen \ea 1993, Schmidt \ea 1997, 
Aguilar Sanch\'ez \ea in prep.) 
from the Palomar-Green catalogue (Green \ea 1986) and from the 
Hamburg Survey (Hagen \ea 1995). 
Other stars have been selected purely because proper motions are available. 

The proper motions used in this work have in part been determined 
by us (see below).  
Further data were taken from
the catalogue of the Lick Northern Proper Motion (NPM) 
programme (Klemola \ea 1987), 
the already published proper motions for subdwarf stars from 
Colin \ea (1994) and from Thejll \ea (1997).

\subsection{New absolute proper motions}

We have determined absolute proper motions for twelve stars from 
recent CCD observations in combination with the Palomar Sky Survey (POSS).
From the sample of stars with known distances we selected several 
which are located in fields with a sufficient number of background galaxies. 
New CCD observations were taken with the 
`Weitwinkel Fl\"achen Photometer and Polarimeter' 
(WWFPP, see Reif \ea 1995) at the 1.2m telescope at the Calar Alto in 1995 
and with the Hoher List Camera (HOLICAM, see Sanner \ea 1997) 
on the 1m telescope at our Hoher List observatory in 1996. 
The positions of the star and the galaxies were used in combination with the 
published APM scans of the POSS as first epoch data.

\begin{table}
\caption[]{Absolute proper motions of stars using 
background galaxies (POSS \& new CDD data)}
\begin{tabular}{lrrrcr}
\hline
Star        & $V$ & \mua    &  \mud & \dmub , \dmud & n\\
\hline
PG 0142+148 & 13.7 & $-$17.4 &  $-$0.4 & 3.7 , 3.9 & 5\\
PHL 1079    & 13.4 &   +11.1 & $-$17.9 & 1.9 , 3.9 & 6\\
PG 0212+148 & 14.5  & $-$3.8 &  $-$9.2 & 1.0 , 1.2 & 22\\
PG 0212+143 & 14.6  &  +11.2 &  $-$1.4 & 1.5 , 1.4 & 13\\
PG 0242+132 & 13.2  &  +17.2 &  $-$9.7 & 1.2 , 1.2 & 31\\
PG 2218+020 & 14.2  &   +1.2 & $-$11.8 & 2.8 , 1.8 & 5\\
PG 2226+094 & 14.1  &  +14.3 &    +0.4 & 1.9 , 2.0 & 7\\
PG 2259+134 & 14.6  &   +0.6 &  $-$9.6 & 1.3 , 2.0 & 12\\
Feige 108   & 13.0  & $-$7.2 & $-$18.4 & 1.4 , 1.4 & 36\\
Feige 109   & 13.8  & $-$3.0 &   +13.2 & 0.9 , 1.2 & 56\\
PG 2337+070 & 13.6 & $-$19.0 & $-$38.8 & 2.4 , 1.7 & 15\\
PG 2349+002 & 13.3 & $-$14.6 & $-$19.3 & 1.6 , 1.7 & 26\\
\hline
\end{tabular}

\noindent
proper motions given in mas/yr\\
\dmub\ = \dmuac\ and \dmud\ are the mean uncertainties of the zero point 
shift of the galaxy positions\\
n = number of galaxies used
\end{table}

The CCD frames were reduced using DAOPHOT to determine
the rectangular coordinates $x$ and $y$. 
The coordinates of the APM scans were transformed to rectangular coordinates 
$x$ and $y$ and a classical astrometric reduction was performed
in a local astrometric system. 
By subtraction of the mean apparent proper motion of the galaxies from those 
of the stars we obtained absolute stellar proper motions. 
The results for our twelve stars are given in Table 2.
This method of getting absolute proper motions was first tested 
in the field of the quasar 3C\,273 (Geffert \ea 1994).

Since these proper motions are based on only one first epoch
position, we are not able to calculate proper motion errors.
An indication of the proper motion uncertainty may be 
the uncertainty of the zero point shift, 
representing a lower limit to our errors.
In Table 2, \dmuac\ and \dmud\ designate the mean uncertainty of the
zero point shift represented by the r.m.s.-values of the apparent 
proper motions of the galaxies.

As best data from the literature we have taken the proper motions 
from the NPM catalogue (Klemola \ea 1987; K+87), 
because these proper motions were calibrated with respect to 
extragalactic objects too. 
In fact, Klemola \ea were among the first to use background galaxies 
to arrive at proper motions in a true inertial system.

In all we have for 21 stars absolute proper motions based on 
extragalactic calibration.

\begin{table*}
\caption[]{Stellar coordinates and orbital characteristics}
	\label{tableofcoordinates}
\begin{tabular}{lrrrrrrrrcrr}
\hline
Name & $l$ & $b$ & $X$ & $Y$ & $Z$ & $U$ & $V$ & $W$  &
    $Iz$  &  $ecc^a$ & $nze^b$\\
      &    &    & kpc & kpc & kpc & \kms & \kms & \kms &
 kpc \kms &  &  \\
\hline 

PG 0004+133  & 106.86 & $-$47.89 &  $-$8.78 &   0.90 &  $-$1.05 &
     75 &   99 &  $-$76 &  $-$941 & 0.50 & 0.34\\
PG 0039+049  & 118.58 & $-$57.63 &  $-$8.77 &  0.49 &  $-$0.89 &
  $-$16 &  214 &  $-$99 & $-$1868 & 0.14 & 0.32\\
CD $-$38 222 & 311.58 & $-$78.86 & $-$8.46 & $-$0.05 & $-$0.32 &
  $-$30 &  200 &     9  & $-$1690 & 0.13 & 0.04\\
PG 0101+039  & 129.10 & $-$58.49 &  $-$8.65 &  0.18 &  $-$0.38 &
   $-$1 &  195 & $-$111 & $-$1687 & 0.04 & 0.36\\
PG 0133+114  & 140.12 & $-$49.71 &  $-$8.88 &  0.32 &  $-$0.59 &
    39 &    72 &      1 &  $-$653 & 0.68 & 0.06\\
PHL 1079     & 144.95 & $-$57.21 &  $-$8.86 &  0.25 &  $-$0.68 &
     9 &   159 &  $-$19 & $-$1410 & 0.27 & 0.09\\
PG 0142+148  & 141.87 & $-$45.79 &  $-$9.14 &  0.50 &  $-$0.84 &
   156 &   237 &    80  & $-$2244 & 0.50 & 0.17\\
PG 0212+148  & 151.20 & $-$43.23 &  $-$9.62 &  0.61 &  $-$1.20 &
    25 &   221 &  $-$86 & $-$2143 & 0.10 & 0.13\\
PG 0212+143  & 151.67 & $-$43.63 &  $-$9.68 &  0.64 &  $-$1.28 &
  $-$103 &  190 & $-$22 & $-$1769 & 0.40 & 0.30\\
PG 0242+132  & 160.87 & $-$40.88 &  $-$9.49 &  0.34 &  $-$0.91 &
  $-$47 &   117 &     9 & $-$1098 & 0.49 & 0.11\\
PG 0856+121  & 216.49 &  33.68 &  $-$9.16 &  $-$0.49 &   0.55 &
  $-$74 &   116 & $-$46 & $-$1099 & 0.48 & 0.14\\
PG 0907+123  & 217.69 &  36.29 &  $-$9.47 &  $-$0.75 &   0.90 &
    $-$6 &   176 &   86 & $-$1675 & 0.12 & 0.31\\
PG 0918+029  & 229.41 &  34.28 &  $-$9.06 &  $-$0.65 &   0.59 &
  $-$46 &  153 & $-$132 & $-$1415 & 0.17 & 0.57\\
PG 0919+273  & 200.46 &  43.88 &  $-$8.74 &  $-$0.09 &   0.24 &
    66 &   213 &     6  & $-$1851 & 0.22 & 0.03\\
PG 1101+249  & 212.76 &  65.87 &  $-$8.64 &  $-$0.09 &   0.36 &
  $-$33 &   256 & $-$56 & $-$2208 & 0.21 & 0.12\\
PG 1114+073  & 250.49 &  59.84 &  $-$8.58 &  $-$0.21 &   0.39 &
     1 &   199 &   $-$9 & $-$1707 & 0.09 & 0.05\\
PG 1232$-$136 & 296.99 & 48.77 &  $-$8.33 &  $-$0.34 &   0.43 &
  $-$79 &   135 &    38 & $-$1152 & 0.42 & 0.09\\
PG 1233+427  & 133.70 &  74.42 &  $-$8.56 &   0.06 &   0.31 &
     17 &   227 &    74 & $-$1942 & 0.09 & 0.18\\
PG 1256+278  &  47.44 &  88.19 &  $-$8.48 &   0.02 &   0.78 &
  $-$71 &   196 &    75 & $-$1666 & 0.26 & 0.21\\
PG 1343$-$101 & 324.21 & 50.15 &  $-$8.13 &  $-$0.27 &   0.55 &
  $-$32 &   149 &    58 & $-$1216 & 0.31 & 0.16\\
PG 1432+004  & 350.07 &  53.31 &  $-$8.05 &  $-$0.08 &   0.61 &
     58 &   139 &    12 & $-$1116 & 0.40 & 0.08\\
PG 1433+239  &  30.56 &  66.35 &  $-$8.34 &   0.10 &   0.43 &
    14 &   189 &  $-$44 & $-$1573 & 0.13 & 0.11\\
PG 1452+198  &  24.66 &  60.82 &  $-$8.14 &   0.16 &   0.71 &
    71 &    169 &    51 & $-$1388 & 0.28 & 0.16\\
PG 1519+640  & 100.27 &  46.17 &  $-$8.58 &   0.44 &   0.47 &
  $-$65 &   360 &  $-$72 & $-$3062 & 0.65 & 0.14\\
PG 1619+522  &  80.47 &  43.96 &  $-$8.41 &   0.55 &   0.53 &
  $-$29 &   211 &  $-$12 & $-$1757 & 0.14 & 0.07\\
PG 1647+252  &  45.14 &  37.13 &  $-$8.10 &   0.40 &   0.43 &
  $-$11 &   265 &     43 & $-$2142 & 0.23 & 0.09\\
PG 1708+602  &  89.28 &  35.91 &  $-$8.48 &   1.45 &   1.05 &
 $-$108 &   164 &     95 & $-$1232 & 0.51 & 0.31\\
PG 1710+490  &  75.43 &  36.08 &  $-$8.35 &   0.56 &   0.42 &
     29 &   216 &  $-$48 & $-$1823 & 0.05 & 0.11\\
PG 1722+286  &  51.71 &  30.57 &  $-$8.04 &   0.59 &   0.44 &
  $-$53 &   211 &      6 & $-$1665 & 0.22 & 0.05\\
PG 1725+252  &  48.21 &  28.73 &  $-$8.11 &   0.43 &   0.32 &
   $-$3 &   233 &     72 & $-$1891 & 0.12 & 0.16\\
PG 1738+505  &  77.54 &  31.78 &  $-$8.32 &   0.80 &   0.51 &
  $-$27 &    241 &    53 & $-$1987 & 0.20 & 0.12\\
HD 205805    & 353.12 & $-$47.81 & $-$8.36 & $-$0.03 &  $-$0.22 &
 $-$100 &   198 &     12 & $-$1656 & 0.32 & 0.03\\
PG 2204+035  &  64.38 & $-$39.83 &  $-$8.11 &   0.82 &  $-$0.76 &
     20 &   261 &  $-$85 & $-$2132 & 0.26 & 0.22\\
PG 2218+020  &  66.05 & $-$43.43 &  $-$8.16 &   0.76 &  $-$0.79 &
     43 &   201 &  $-$41 & $-$1676 & 0.08 & 0.14\\
PG 2226+094  &  74.78 & $-$39.58 &  $-$8.27 &   0.84 &  $-$0.72 &
  $-$49 &   241 &  $-$49 & $-$1954 & 0.26 & 0.12\\
PG 2259+134  &  86.36 & $-$41.31 &  $-$8.43 &   1.03 &  $-$0.91 &
     39 &   209 &  $-$44 & $-$1803 & 0.03 & 0.16\\
Feige 108    &  76.82 & $-$55.94 &  $-$8.45 &   0.21 &  $-$0.33 &
     43 &   235 &  $-$35 & $-$1996 & 0.15 & 0.07\\
Feige 109    &  86.55 & $-$48.25 &  $-$8.45 &   0.75 &  $-$0.84 &
  $-$14 &   241 &     37 & $-$2026 & 0.16 & 0.13\\
PG 2337+070  &  93.71 & $-$51.49 &  $-$8.53 &   0.48 &  $-$0.60 &
    140 &   150 &  $-$31 & $-$1351 & 0.47 & 0.10\\
PG 2349+002  &  93.10 & $-$58.91 &  $-$8.52 &   0.42 &  $-$0.70 &
     98 &   160 &     58 & $-$1407 & 0.37 & 0.15\\
PG 2358+107  & 103.53 & $-$49.96 &  $-$8.62 &   0.52 &  $-$0.64 &
     49 &   193 &   $-$9 & $-$1691 & 0.15 & 0.08\\
\hline
\end{tabular}

$^a$ $ecc$ = eccentricity of the orbit, $(R_a - R_p)/(R_a + R_p)$\\
$^b$ $nze$ = normalised $z$ extent of the orbit, 
$z_{\rm max}/(\varpi\,\, {\rm at}\,\, z_{\rm  max})$
\end{table*}

As second priority data we have taken the proper motions of 
Colin \ea (1994; C+94). 
The proper motions of this set have the highest internal 
accuracy (1-2 mas/yr). 
Since these data are based on the PPM catalogue (R\"oser \& Bastian 1991) 
we may expect additional systematic errors of our proper motions 
due to the local inhomogeneties of the PPM catalogue. 
A full use of the internal accuracy of these data will eventually be 
possible through a rereduction with Hipparcos reference stars. 

Finally, proper motions were determined by Thejll \ea (1997; T+97) 
from recent accurate meridian observations and the published positions 
of the Astrographic Catalogue. 
Since they did not sufficiently describe from which reduction the old 
position was taken, and 
since these proper motions are based on only one or two first epoch plates, 
we used these proper motions with least priority.

\begin{figure*}
\def\epsfsize#1#2{1.0\hsize}
\centerline{\epsffile{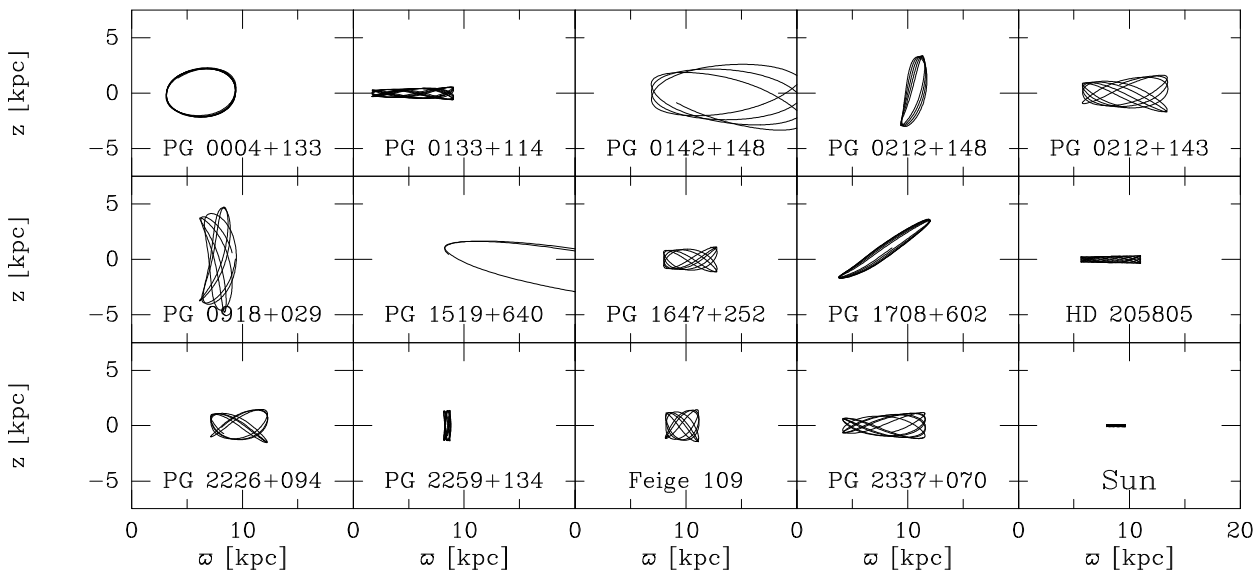}}
\caption[]{For several stars the orbits are shown to demonstrate 
the variety in shape. 
The diagram shows the meridional cut, i.e., the plane through 
the rotation axis of the galaxy rotating along with the motion of the star. 
Plotted is the motion of the star in that plane 
in vertical distance $z$ and galactocentric distance $\varpi$. 
All orbits were calculated backward over 1 Gyr in steps of 1 Myr. 
For comparison we have added the orbit of the Sun}
\end{figure*}

For the stars common to these samples we have compared the proper motions. 
There are 11 stars in the list of T+97 common with the NPM (K+87). 
The mean of the differences with their r.m.s. deviations are 
$+0.2 \pm$ 6.0 mas/yr in \mua\ and 
$+1.7 \pm$ 8.3 mas/yr in \mud\ 
(in the sense NPM minus T+97). 
For the five stars common to the catalogues by Colin \ea (1994) and 
Thejll \ea (1997) we  found mean differences 
(in the sense T+97 minus C+94) 
of $-5.6 \pm$ 3.8 mas/yr in \mua\ and $+2.6 \pm$ 2.3 in \mud. 
Since the error of a single proper motion in the NPM is of the order 
of 5 mas/yr, we expect from our comparison nearly the same accuracy 
for the proper motions of T+97. 
The better agreement between the catalogues of C+94 and T+97 may 
be explained by the fact that the local inhomogenities
of the PPM catalogue will affect both methods in a similar way.

\subsection{Radial velocities}

Radial velocities are available for the stars from the spectra 
in the Bonn data base (see papers cited with Table 1). 
Several of them have already been published, in some cases 
they have been determined for this paper. 
For stars from S+94 radial velocities have been taken from Saffer (1994). 
The radial velocities have accuracies of the order of 30 \kms. 

For some stars it is known (see Table\,1) 
that the radial velocities are variable. 
This may be a sign for binary nature of the objects 
(see, e.g., Theissen et al. 1995, Paper V). 
However, the physical parameters derived for the selected stars 
were beyond doubt, so that the distance is reliable. 
We have investigated the effect of changes in the radial velocities 
on the orbits, as described in Sect.\,5.2, 
and found them only of minor importance in a statistical sense.

\subsection{Distances}

The distances of the stars have been taken from the literature, 
as indicated with Table 1.
Distances are based on the determination of \teff\ and \lgg, 
the reddening corrected visual magnitude, 
and the assumption that the star has a mass of 0.5 \ms. 
The distances are accurate to about 30\%. 
For a discussion of the distance determination method 
see the papers cited with Table 1.

For 2 stars distances are based on Hipparcos parallaxes 
from de Boer et al.\, (1997; B+97).

\section{Orbits, populations}

\subsection{The mass model for the galaxy}

In order to calculate orbits a model for the gravitational potential of
the Milky Way has to be adopted. 
We have based our study on the model potential by Allen \& Santillan (1991). 
This model was particularly developed for use in an orbit program 
and has been applied to the orbits of nearby stars, metal poor stars, 
as well as globular cluster orbits (Allen \& Santillan 1991, 1993, 
Scholz et al. 1996, Schuster \& Allen 1997).

An alternative model of the kind which is suitable for numerical orbit 
integrations would, for instance, be the one of Dauphole \& Colin (1995).
However, previous investigations have shown that the results obtained 
with these different models agree as long as the orbits do not reach
extreme distances from the galactic centre (Dauphole et al. 1996).

We used the Allen \& Santillan model as implemented in an updated version 
of the computer program of Odenkirchen \& Brosche (1992). 
In order to be consistent with the parameters of the model, 
our calculations of the stellar space velocities follow the current IAU 
values for the LSR $\Theta_{\rm LSR}$ = 220 \kms\ and $R_{\rm LSR}=8.5$ kpc.

\subsection{Calculated orbits}

The observational data (Table 1) have been transformed into the 
positional ($X,Y,Z$) and velocity ($U,V,W$) coordinates 
in the galactic system (Table 3). 
Note that the $X$-axis points from the Sun toward the galactic centre 
with the zero point at the galactic centre.

The orbits were calculated over a total of 1 Gyr backwards. 
This time span is longer than the sdB evolutionary phase 
(assuming they are genuine horizontal-branch like stars), 
but we opted to use 1 Gyr to achieve good statistics for the 
stellar positions (see Sect.\, 4). 

A selection of the orbits is shown in Fig.\,1. 
Given are the meridional cuts, showing the motion of the star in 
projected galactocentric distance \w\ and in distance to the plane 
of the Milky Way. 
The figure demonstrates the variety of orbits found. 

The orbits of the stars of our sample are rather well behaved. 
Most stars stay overall very close to the disk but 
10 stars veer out to $z>2$ kpc (PG 0918+029 reaches $z \simeq 5$ kpc). 
Many of the stars have orbits reaching way in toward the centre 
of the Milky Way (7 stars to $\varpi <4$ kpc, 
the extreme is PG 0133+114 to $\varpi \simeq 1.8$ kpc). 
8 stars move out to $\varpi > 12$ kpc. 
A most notable result is that the orbits cover large portions of 
the galaxy (see also Fig.\,5).

\subsection{Disk and Halo orbits, populations}

We have attempted to sort the stars according to `halo'-like and 
`disk'-like orbits. 
When trying to do so one has to consider the formation history of the 
stellar populations in our galaxy. 

The stars of the globular clusters were among the first to be formed 
in the galaxy; they are called Population II stars. 
Relatively recently formed stars, like those of the open clusters, 
are part of the Population I. 
The very different morphology of the data point distribution in the respective 
colour-magnitude diagrams led Baade to the concept of these two populations. 
Star formation has most likely been a continuous process in our galaxy. 
This must mean that there is, in practice, 
a population continuum without sharp boundary between these populations, 
a fact exemplified in phrasings like old population I, 
old disk population, and the like. 

In principle the sdB stars in the galaxy form a mix of stars of 
a large age range. 
Stars having started with about 2.5 \ms\ evolve in about 0.3 Gyr 
to the horizontal-branch stage, 
whereas old stars having started as main-sequence star with $\sim 0.8$ \ms\ 
take some 12 to 15 Gyr to become the sdB star it is today. 
With a constant star formation rate in the galaxy 
we would expect that the sdB stars of today come from the past
in proportion to the initial mass function. 
So we would expect that sdB star samples are dominated by the older ones. 

The kinematics of the formation location may be reflected in the 
motion of today. 
Stars having formed in the halo still will have `halo' orbits, 
while stars having formed in the galactic disk 
are expected to have orbits confined to the disk. 
However, dynamically interacting encounters with other stars during 
the full stellar life time will have had its effect on the kinematics too. 
For that reason one may expect that old stars of the disk
have heated-up orbits.

\begin{figure*}
\def\epsfsize#1#2{1.0\hsize}
\centerline{\epsffile{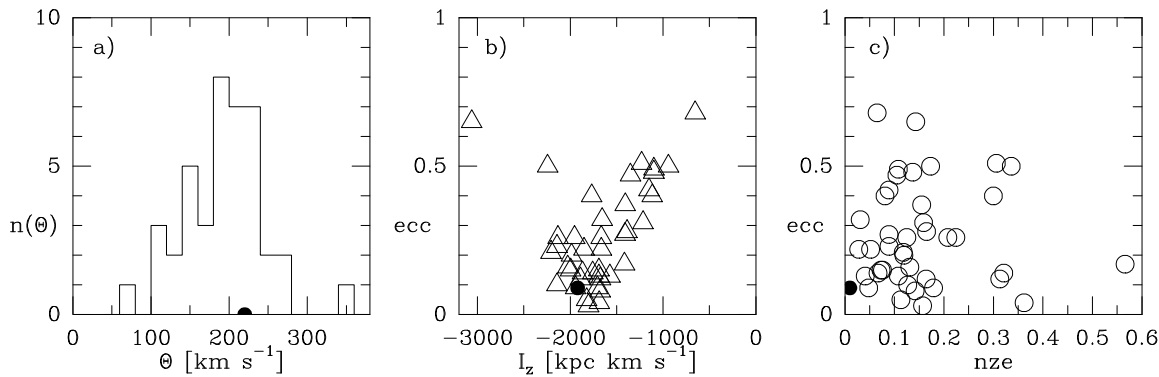}}
\caption[]{Several parameters of the stellar orbits are plotted 
(for a discussion see Sect.\,3.3). 
Panel a) shows the histogram of the values of 
the present velocity component $\Theta$. 
Panel b) shows the orbit parameters eccentricity, 
$ecc = (R_a - R_p)/(R_a + R_p)$, 
and angular momentum, $I_z$. 
Note the large number of stars with highly eccentric orbits 
as well as the stars with Sun like orbit parameters (see Sect.\,3.3).
Panel c) shows the values for the eccentricity, $ecc$, 
together with the normalised $z$ extent, 
$nze = z_{\rm max}/(\varpi\,\, {\rm at}\,\, z_{\rm max})$, 
or the halo-ness of the orbit. 
In all panels the value for the Sun is indicated as $\bullet$}
\end{figure*}

We have calculated several parameters related to the orbits of the stars 
(see Table 3). 
$\Theta$, the velocity component parallel to the circular galactic rotation 
(cylindrical coordinates) 
shows a distribution with a maximum centered near the solar 220 \kms\ 
but with a rather large spread to $\simeq 100$ \kms\ (see Fig.\,2a). 
For the angular momentum $I_z$ the same holds, with a peak in the 
distribution at $I_z = -1700$ kpc \kms\ 
in particular spreading to $-700$ kpc \kms\ (see Fig.\,2b). 
Statistically one may expect a spread in each orthogonal velo\-city component 
of about 30 \kms, based on (on average) errors in the radial velocity 
as well as in each component of the proper motion of about 30 \kms. 
The observed spread in $\Theta$ (and $I_z$) is much larger than that 
and thus it is not due to noise in the input data. 
The stars with small $\Theta$ (small $|I_z|$) 
most likely are the older ones in the sample. 
Note that the peak  $I_z = -1700$ kpc \kms\ corresponds to 
$\Theta = 197$ \kms, less than the solar value.

Another parameter of relevance is the eccentricity of the orbit 
defined as $ecc = (R_a - R_p) / (R_a + R_p)$, 
with $R_a$ and $R_p$ the apo- and perigalactic distances of the stars 
(as in Allen \ea 1991). 
The average for our stars is 0.24 but a notable number has $ecc>0.4$ 
(note that the orbit of the Sun has $ecc=0.09$ in the above definition). 
The orbits of stars with small $|I_z|$ have, of course, 
large eccentricities (see Fig.\,2b). 

The orbits show the effect of diminishing gravitational potential 
with galactocentric distance: 
each star moves to larger $z$ when at larger \w\ 
than at smaller projected galactocentric distance. 
In order to properly quantify the maximum height the star reaches 
outside the disk we have defined the `normalised $z$-extent', 
$nze = z_{\rm max}/(\varpi\,\, {\rm at}\,\, z_{\rm max})$. 
The value of this parameter for each orbit is given in Table 3.
The average for our sample is 0.16. 
A large value of $nze$ is an indication for a halo-like orbit. 

The parameters $ecc$ and $nze$ are plotted together in Fig.\,2c.
In general one would expect that orbits very different from that of 
the Sun would be those of old stars. 
We therefore suspect that the stars with orbits with approximately either 
$I_z \geq -1400$ kpc \kms, or $ecc \geq 0.25$, or $nze \geq 0.25$, 
in general be considered to be the older ones. 

Can we identify individual stars as old ones based on their orbit?
PG 0918+029 reaches to $z= 5$ kpc, the largest $z$-value in the sample, 
while PG 1519+640 has a very elongated orbit reaching $\varpi = 20$ kpc. 
These two stars do not have, however, proper motions based on an 
extragalactic reference. 
PG 0212+148 reaches to $z \simeq 3.5$ kpc 
(the $z$-extent of this orbit is much more limited than the one given 
by Colin et al., 
essentially due to a more accurate proper motion). 
PG\,0142+148 has an orbit with \w\ covering the range 
of $7 \leq \varpi \leq 22$ kpc. 
These orbits may be the halo like ones, 
but none of the stars of our sample exhibits clear halo orbit characteristics. 

Finally, the average kinematic properties of our sample may also be of 
relevance for characterising the sdB star population. 
We have calculated the mean asymmetric drift which turns out to be 
$-36\pm 7$ \kms. 
The dispersion in the values of the kinematical parameters is 
$\sigma_{\rm U} = 62 \pm 8$ \kms, $\sigma_{\rm V} = 52 \pm 7$ \kms, and 
$\sigma_{\rm W} = 59 \pm 8$ \kms. 
These values are in very good agreement with the kinematical properties 
of thick disc stars (Ojha \ea 1994, and references therein).

\subsection{Discussion of the results}

We conclude that our sample does not contain stars 
which can be uniquely identified as old and thus as Population II stars. 
Either the sample is still too small 
or over time all orbits have been modified to general disk like orbits. 
However, several stars have orbits indicative of larger age, 
identified in Fig.\, 2b and Fig.\,2c as those 
whose orbital parameters are very different from the Suns orbit parameters. 

The orbits found are generally well behaved and there are no orbits 
indicating chaotic behaviour. 
Schuster \& Allen (1997) have investigated a large sample of 
metal poor high-velocity stars. 
These do show chaotic orbits, but Schuster \& Allen conclude this shows up 
predominantly in stars whose orbits reach to 
galactocentric distances \w $\leq 1$ kpc. 
Our sdB stars stay all further out. 
The Schuster \& Allen star orbits also reach to much larger $z$-values 
than those of our sdB stars. 

The relatively large values of $\Theta$ (and of $|I_z|$) for our orbits 
suggest that, as a sample, the sdB stars move in the galaxy 
not too dissimilar from the LSR. 
In fact, the average properties of the orbits are quite different from 
those of the metal poor high-velocity stars studied 
by Allen \ea (1991) and Schuster \& Allen (1997). 
This difference may be an indication for a difference in origin. 
The sdB stars either are not very metal poor and not very old 
while the metal poor high-velocity stars are much older. 
However, if those metal poor stars are older than the sdB stars, 
one wonders where the horizontal-branch like stars 
emerging from such an old population have gone. 
Alternatively one could speculate about a completely 
different origin for the sdB stars (see e.g. Paper V).
More stars have to be investigated to clarify these questions.

\section{$z$-Probability and $z$-distribution}

\subsection{$z$-Probability in the orbit}

We have investigated the probability with which one may find a given star 
at a particular $z$-distance. 
For that we calculated the orbits with equal time steps of 1 Myr. 
The statistics of these $z$-values gives the frequency $N(z)$ to 
find the star at a given $z$. 
Examples of such histograms are shown in Fig.\,3.

\begin{figure}
\def\epsfsize#1#2{1.0\hsize}
\centerline{\epsffile{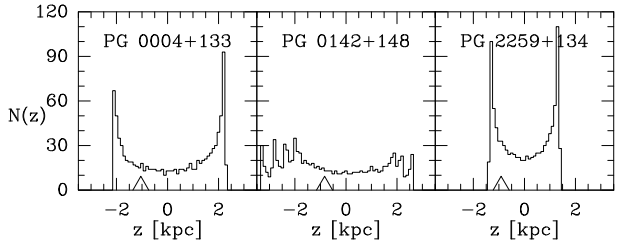}}
\caption[]{Sample histograms showing the frequency $N(z)$ with which a star 
is found at a given height in $z$ 
when considering fixed time intervals in the orbits calculated over 1 Gyr. 
The histogram intervals are $z$= 100 pc. 
The orbits themselves are included in Fig.\,1. 
The present location of each star is indicated with $\Delta$ next to 
the $z$-axis}
\end{figure}

The frequency $N(z)$  has a relative minimum near $z = 0$ kpc 
while their maxima are away from the disk. 
This is to be expected. 
All stars will spend more time in the parts of the orbit away 
from the centre of the gravitational potential 
(where the star is far from the disk and the galactic centre) 
with a relatively small space velocity, 
compared to the larger $z$-velocity when the star is nearer 
to the centre of the gravitational potential. 
This means that on average a star spends more time away from the disk 
than `in' the disk. 
However, selection effects may play a role (see Sect.\,4.3.1).

\subsection{$z$-Distribution from the entire sample}

Working with a large sample of stars, we can add the histograms together 
resulting in the overall probability for such stars. 
This statistical distribution in $z$ is shown in Fig.\,4.
In fact, the histogram shows the average spatial distribution 
of such stars, based on their actual kinematic behaviour. 

The overall $z$-distribution for the stars is rather smooth. 
Fitting an exponential to the data one finds a scale height $h_z$= 0.97 kpc, 
based on the combination of the $+z$ and the $-z$ side of the histogram. 
The one-sided values are $h_{z+}$=0.85 kpc and $h_{z-}$=1.05 kpc, 
suggesting that the uncertainty in the derived overall scale height is 
of the order of 0.10 kpc.

The histogram is not identical to an exponential distribution in $z$ 
(Fig.\,4c). 
However, assuming an exponential and determining the slope of ln $N(z)$ 
allows to characterize the distribution with one number, 
which is of great help for the tests and comparisons to be performed. 

One has to note that toward high $z$ the distribution becomes biased 
to the very few stars reaching that far in $z$. 
In fact, overlooking all orbits, no star in the present sample 
reaches further than 6 kpc, while just 3 stars reach distances 
up to 3 to 6 kpc. 
The statistics in this $z$-range is therefore one of small numbers 
and cannot reliably be used for further numerical analysis. 
At the same time we concluded in Sect.\,3.4, 
that the overall sample contains just very few stars (if any) 
stars going to Population II like $z$. 

At small $z$ the distribution found (Fig.\,4 panel b) 
clearly differs from an exponential space distribution. 
The overall $N(z)$ shows a relative minimum near $z$ = 0 kpc, 
i.e., it is more likely to find such stars 
at some distance {\it away} from the disk than {\it in} the disk. 
This finding has consequences for the concept of scale-height fitting 
from `statistically complete samples' in a given direction, 
as used in several investigations. 
Looking back to studies of sdB star scale heights from stellar distances, 
it is clear that the distribution has to be sampled 
to well beyond 1 kpc in $z$ 
to avoid problems with the relative minimum in the real spatial distribution. 

We have calculated the orbits over 1 Gyr, 
although the phase life of sdB stars is about 10$^8$ yr. 
Doing the statistics for just that part of the orbit results 
in a scale height of 0.98 kpc, basically the same as our main value. 
Having used 1 Gyr apparently does not influence the scale height. 

The scale height derived from $N(z)$ is based on all positions in all orbits. 
However, the orbits cover large portions of the galactic plane. 
As noted in Sec.\,3.3, the change in galactic potential with 
projected galactocentric distance \w\ leads to a `thickening' of the orbit. 
We therefore have redone the statistics in three intervals of \w. 
For $2 <$ \w $< 7$ kpc we find $h= 0.73$ kpc,
for $7 <$ \w $< 10$ kpc (the solar vicinity) we find $h= 0.88$ kpc,
for $10 <$ \w $< 20$ kpc we find $h= 1.58$ kpc. 
We conclude that the value from the full sample using all parts 
of the orbits is somewhat biased toward the large \w\ portions. 
The value for the scale height of the sdB stars in the solar 
vicinity is therefore $0.9\pm 0.1$ kpc. 

Before rushing to conclusions we will test in the next subsection 
the robustness of the result against variations in the input parameters. 
It will be shown that small adjustments in the final value 
of the scale height are needed.

\begin{figure*}
\def\epsfsize#1#2{1.0\hsize}
\centerline{\epsffile{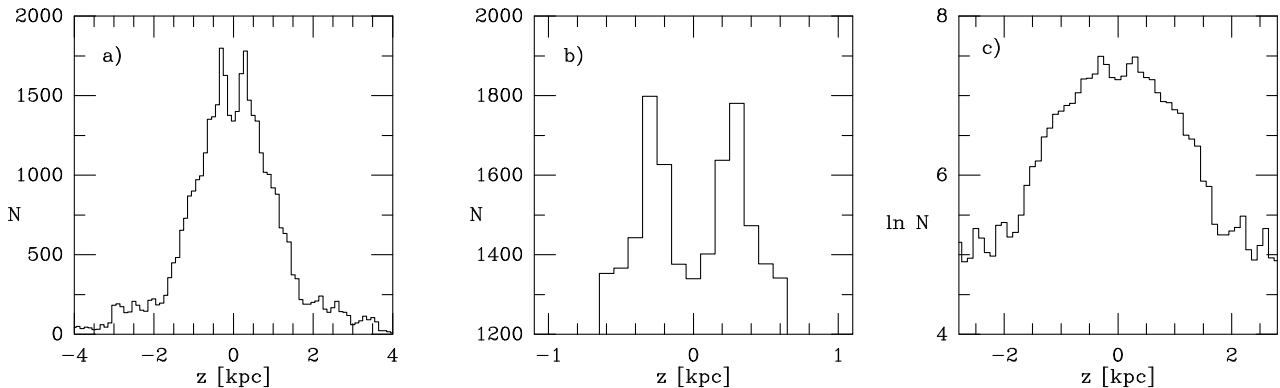}}
\caption[]{Plot of the relative frequency $N(z)$, 
based on all stars of our sample, 
to find a star at a given 
$z$-distance in the Milky Way (histogram bins are 100 pc wide). 
Note that for $|z| \leq 2$ kpc the histogram is based on rapidly 
decreasing numbers of stars and the details of its shape there 
are thus not of significance. 
Panel a): Overall distribution (linear scales). 
Panel b): Enlarged plot (linear scales) of the frequency out to 1 kpc; 
note the rise and fall of $N$ within this range.
Panel c): Logarithmic plot of ln $N(z)$ vs. $z$, showing that 
the distribution is consistent with an exponential 
one with a scale height of $\simeq 1$ kpc}
\end{figure*}

\subsection{Discussion of sources of error}

\subsubsection{Selection effects?}

One may be concerned that selection effects have played a role 
in arriving at our results. 
Let us consider the ways in which the sample came together. 

First, for all stars distances and radial velocities must be available. 
The selection of the stars (from the PG) for the investigations 
of Papers II..VII was essentially random on the sky 
so that no preference for any direction in the galaxy is present. 
However, our data taking started generally with the brighter stars. 
It means that the stars are on average relatively close by. 
But, for each star this proximity is only at the present epoch 
and we therefore sample these individuals by chance (see Fig.\,5). 

On the other hand, for the PG one has not attempted to survey the 
low galactic latitude portion of the sky 
(since the PG was aimed at finding quasars) 
and it does not cover the southern sky. 
In all, the sample therefore lacks stars in some directions, 
as visible in Fig\,5.
It may therefore be possible, that our sample underrepresents stars 
with orbits staying always very close to the disk 
(like that of the Sun; see Fig.\,1). 
Adding such stars might fill in the relative minimum in $N(z)$ at $z$=0 kpc. 

Secondly, good proper motions can be determined for stars which 
have ample first epoch data. 
Since fields of the old plates are normally not defined in terms 
of galactic coordinates but based on the equatorial system, 
the first epoch aspect does not introduce a galactic bias. 
This is, e.g., true for the stars lying in the Bordeaux Zone of 
the {\it Carte du Ciel}.
Yet, the low limiting magnitude of available first epoch plates 
biases our sample to the brighter and thus nearer ones. 
Especially, the proper motions of the objects in the list of T+97 
are based on the Astrographic Catalogue 
which is limited to stars brighter than $m_V \simeq 11$ mag.
Therefore, the T+94 list has only the brightest sdB stars. 
On the other hand, the proper motions determined using the POSS as first 
epoch data (Sect.\,2.1) 
pose in principle not a significant limit in terms of brightness. 
The sample of the Lick stars is not defined in magnitude range, 
since K+87 selected stars of all magnitudes being of astrophysical relevance 
at that time, even as faint as $V= 18$ mag.

\subsubsection{Robustness of the scale height value}

In order to verify that our results do not depend in a critical way 
on the input parameters used, 
we have experimented with the input data for the orbit calculation. 
As indicated in Sec.\,4.2, we will assume that the histogram $N(z)$ 
can be represented by an exponential, for easy comparison. 
We made 3 kinds of experiments. 

$\bullet$ Dividing the sample in two parts

For the half sample with stars now at $90^{\circ} < l < 270^{\circ}$ the 
scale height from the total orbits was 1.30 kpc 
whereas for the stars now being in the interior galactic half 
the scale height came out at 0.72 kpc. 
The average is again $\simeq 1.0$ kpc, 
and the difference reflects the difference in scale height 
in relation with $z/$\w.

\begin{figure}
\def\epsfsize#1#2{1.0\hsize}
\centerline{\epsffile{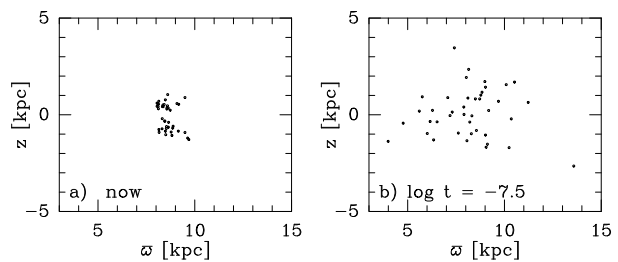}}
\caption[]{Location of the stars in space, projected onto 
the meridional plane ($z$, \w). 
Panel a) The present location is shown. 
The absence of stars in the general 
direction of the plane of the Milky Way is apparent (limits of the PG catalog).
Panel b) Location of the stars 3$\times 10^7$ yr ago 
(being half the time of the sdB evolutionary phase), 
showing that the stars observed near the Sun came from 
a large variety of positions in the galaxy}
\end{figure}

For the half sample with stars now at $|b|> 45^{\circ}$ the full orbits 
gave the scale height 1.0 kpc, 
the stars now being at $|b|< 45^{\circ}$ led to a scale height of 1.13 kpc.

$\bullet$ Variation of radial velocity and proper motion

We added 30 \kms\ to all radial velocities 
(being the observational uncertainty), 
repeated the orbit calculation, 
and made the $z$-distribution statistics. 
In a second attempt we reduced all radial velocities by 30 \kms. 
We found from these experiments the values $h_+= 1.22$ and $h_-= 1.15$ kpc. 
Both values are larger than our original one, 
suggesting that the added error makes the average result less reliable.
The actual $v_{\rm rad}$ gives the smallest scale height. 

We also added 5 mas/yr both in \mua\ and \mud\ and redid our calculations. 
We now found $h=$ 1.04 kpc, within the uncertainty range of our original value.

We conclude that, given the size of the star sample, 
random errors in the input radial velocities and proper motions 
do not affect the value of the scale height in an essential way. 

As a last test here we calculated the histogram based only on 
the orbits of the 21 stars which have {\it absolute} proper motions 
(from Table\,2 and from K+87). 
In this case $h$ comes out at 0.84 kpc and the remainder of the star 
orbits lead to a scale height well above 1 kpc. 
Possibly the C+94 and T+97 proper motions may be affected by 
additional systematic errors which lead to an increase 
in the $z$-distance in the orbits. 
This effect is similar to that of changing the radial velocities. 

$\bullet$ Different distance scale

One of the input parameters is the stellar distance. 
Distance values have uncertainties of the order of 30\%. 
We have not varied the input distances but tested the effects of 
distance errors in the following manner. 

In the research on sdB stars a systematic difference exists 
between values of $T_{\rm eff}$ and log $g$ derived by some groups 
(Saffer \ea 1994) and by our group (Papers II,IV,VII). 
This difference leads to different distances for the stars 
(factors of 1.5 smaller distances from S+94 are not uncommon). 
In our orbit sample we have included stars investigated by both groups. 
We therefore divided our sample in two, one part using our distances and the 
other part using distances derived by S+94. 
For both groups the orbits were calculated and the $z$-distribution 
was determined. 
For the 32 stars from our data we find $h_z$= 1.07 kpc, 
while for the 17 stars with Saffer \ea distances (there is some overlap) 
we find a scale height of 0.76 kpc. 
Changing the distance in a systematic way does make a difference 
(it changes also the tangential velocities). 

$\bullet$ Different gravitational potential

The value of the scale height found is, of course, 
also a function of the nature of the potential model for the galaxy. 
If a smaller surface mass density is assumed, 
the vertical force will be smaller and consequently the scale height larger 
(see Allen \& Santillan 1991). 
For the present study we will not explore these possibilities further. 

\subsection{Final scale height value and discussion}

Overlooking all the tests, we conclude to the following for the 
$z$-distribution of the sdB stars. 
The all-orbit $N(z)$ is consistent with an exponential distribution 
with scale height $h_z$= 0.97 kpc. 
This scale height turned out to be biased somewhat to the $z$-values 
of stars reaching to large \w, 
because $N(z)$ in just the solar vicinity indicates that the 
base value is to be reduced to $h_z$=0.88 kpc. 
Subsets of the sample gave essentially the same scale height as the 
base value of 0.97 kpc, with the just noted exception of the division 
in inner galactic and outer galactic stars.
Variations in the input velocities did not produce dramatic changes 
in the base value. 
Changing the distance of the stars in a systematic manner  
(tested by using the distances from Papers II,IV,VII versus those from S+94) 
lead to a difference in scale height of a factor 1.4. 
Taking the S+94 stars out of our data set means increasing the 
base value from 0.97 to 1.07 kpc

Combining these results from the tests 
we conclude that the sdB stars scale height $h_z$ is best represented 
with the value of $1.0\pm 0.1$ kpc. 

The scale height derived from our orbit data 
is much larger than the $\simeq200$ pc derived for the sdB stars 
by Heber (1986) and in Paper II and IV. 
Clearly, such studies do not sample sdB stars to large enough distances, 
or they undersample the number of stars close by. 
It is well known that the value of a scale height is largely determined 
by the extreme points in such a distribution 
and both ends have a large risk of being unreliable given the 
small number of stars in those extreme bins. 
Also, the relative maximum near $z=300$ pc in $N(z)$ of Fig.\,4 
makes clear that samples over limited distances can by definition not 
lead to a good characterisation of the true $z$-distribution. 

In an analysis of the spatial distribution of sdB stars 
Villeneuve \ea (1995) derived the stellar temperatures from photometry, 
the garavity by using a fixed relation between \teff\ and log $g$ from 
Greenstein \& Sargent (1974), and thus could calculate distances. 
Since they used just photometry, 
instead of going through a more detailed spectroscopic analysis, 
substantially larger distances could be reached. 
Villeneuve \ea (1995) then found indications for scale heights 
ranging from 500 to 900 pc. 
That range tends more to what we have derived based on the orbit statistics. 

The calculated mean asymmetric drift of $-36$ \kms\ points, 
together with the scale height of about 1\,kpc, 
to a population of thick disc stars. 
These parameters are in remarkable agreement with the studies of 
Ojha et al. (1994) and Soubiran (1993). 
Those studies are based on investigations in limited fields 
(5\degr $\times$ 5\degr),
whereas our study used a comparatively small number of objects 
but distributed over a large area of the sky.

\section{Conclusions}

The analysis of the galactic orbits of 41 stars shows that the 
frequency distribution of finding these stars at a given $z$ 
can be given by an exponential in $z$ 
with a scale height of 1.0$\pm$0.1 kpc. 
Our results are, as various tests with variations of the input 
parameters have shown, robust and reliable with the indicated error.
That value is in agreement with the notion that sdB stars are part 
of the older disk population. 
The asymmetric drift value supports this conclusion. 

Since sdB stars most likely are the end product of 
evolution of stars with mass ranging from 0.8 to at most 3 \ms, 
stars of different age are present in the sample. 
That evolutionary time from the zero age main-sequence to sdB state 
ranges from 0.5 Gyr to over 12 Gyr. 
In that time the number of gravitational interactions with other stars 
has apparently been large enough to erase most traces of their origin. 

Our sample contains 10 stars (one out of four) 
with orbits reaching to $z > 2$ kpc, 
the extreme being the star reaching 5 kpc. 
Several stars have highly eccentric orbits, a small angular momentum $I_z$, 
or orbits with a large normalised $z$-extent. 
These all are likely the older stars in the sample.

\acknowledgements
We thank Ralf Kohley (Bonn) and the staff of the Calar Alto Observatory 
for their help in obtaining the WWFPP data at the 1.23m telescope 
and Klaus Reif (Bonn) for his support with the Hoher List observations. 
We thank the referee, Dr.\,O.\,Bienaym\'e, for suggesting to include a 
calculation of the asymmetric drift. 
This research project was supported in part by 
the Deutsche Forschungs-Gemeinschaft under grant Bo 779/11 
as well as through observing support grants Bo 779/12 and Bo 779/18. 
YAS thanks the DFG for a doctoral thesis stipendium in the framework of 
cooperation between the DFG and the Instituto de Astrof\'{i}sica de Canarias. 
MO was supported by a grant from the 
Bundesministerium f\"ur Forschung und Technologie (FKZ 0100023 6) and 
by the European Community through a Marie Curie research fellowship 
(ERBFMBICT 961511).
For our research we made with pleasure use of the SIMBAD in Strasbourg.

\end{document}